\documentclass[12pt]{article}
\usepackage{graphicx,amsmath} % Required for inserting images

\usepackage{times,graphicx,parskip,natbib}
\usepackage{amsmath,setspace}
\usepackage{amssymb}
\usepackage{accents}
\usepackage{float}
\usepackage{subfigure,authblk}
\usepackage[colorlinks=true,allcolors=blue]{hyperref}

\begin{document}

\title{Compositional regression using principal nested spheres}
\author[1,3]{Mymuna Monem}
\author[2]{Ian L.~Dryden\footnote{Corresponding author email: ian.dryden@nottingham.ac.uk}}
\author[1]{Florence George}
\author[4]{Natalia Soares Quinete}

\affil[1]{Department of Mathematics and Statistics, Florida International University, Miami, FL, USA} 
\affil[2]{School of Mathematical Sciences, University of Nottingham, Nottingham, UK}
\affil[3]{Baptist Health South Florida, Coral Gables, FL, USA} 
\affil[4]{Department of Chemistry and Biochemistry, Institute
of Environment, Florida International University, North Miami, FL, USA.}

\date{ }

\onehalfspacing

\maketitle

\begin{abstract}
Regression with compositional responses is challenging due to the nonlinear geometry of the simplex and the limitations of Euclidean methods. We propose a regression framework for manifold-valued data based on mappings to statistically tractable intermediate spaces. For compositional data, responses are embedded in the positive orthant of the sphere and analysed using Principal Nested Spheres (PNS), yielding a cylindrical intermediate space with a circular leading score and Euclidean higher-order scores. Regression is performed in this intermediate space and fitted values are mapped back to the simplex. 
A simulation study demonstrates good performance of PNS-based regression. 
An application to environmental chemical exposure data illustrates the interpretability and practical utility of the method.

\end{abstract}

%v4 sent off to arXiv
%v5 sent off to Stat journal
\section{Introduction}
Compositional data arise in a wide variety of fields, 
where there is a vector of measurements which are all non-negative and the components sum to a constant \citep{Aitchison86}. A typical situation is where the data are proportions in each of 
a set of classes and the proportions sum to 1. 
Examples include the proportions of voters voting for each political party in an election; the proportions of chemical elements in an environmental sample; and the proportions of assets in different types of stocks in a financial portfolio. 
In compositional data analysis, it is the relative sizes of each component that are important, rather than the absolute sizes. 
Statistical analysis must be 
adapted because of the constant sum constraint. Compositional data analysis was introduced to a wide statistical audience by \cite{Aitchison86}, and a broad overview 
of the field is given by \cite{Boogtolo13}.

Compositional data are a form of object data, and an introduction to 
the framework of object oriented data analysis is given by \citet{Marrdryd22}. Common tasks include defining a mean object, exploring the structure of object variability, dimension reduction, and carrying out regression with predictors and responses in different types of object spaces. 
Regression methods for compositional data have traditionally been based on log-ratio transformations \citep{Aitchbaco84} or hyperspherical representations \citep{Scealwels11}, while regression on manifolds more generally is often carried out via intermediate or tangent-space constructions \citep{Fletcher13,Pennec06}.

There are many choices for transforming compositional data in
order to carry out statistical analysis. The presence of zeros can cause problems, and this may be missing information, actual zeros or values below a detection limit. Using log-transformations 
is particularly problematic in this case. An alternative 
is to transform the composition to a sphere using a square root transformation \citep{Scealwels11}, and then make use of the techniques of spherical data analysis. We shall adopt this approach in this paper. 

\section{Compositional data}
\subsection{Transformation of compositional data to a sphere}
Suppose that we have $d+1$ multivariate non-negative components  $x_j \ge 0, j=1,\ldots,d+1$, and we make the data compositional by normalizing to satisfy the unit-sum constraint   $\sum_{j=1}^{d+1} x_j = 1$. The sample space of such compositional data is a $d$-dimensional simplex given by\\ 
$${\displaystyle \Delta^{d}=\left\{\mathbf {x} = (x_{1},x_{2},\dots ,x_{d+1})^T \in \mathbb {R} ^{d+1}\,\left|\,x_{j}\geq 0,j=1,\dots ,d+1;\sum _{j=1}^{d+1}x_{j}=1 \right.\right\}.\ }$$ 

For our analysis we will use a transformation of the compositional data to the positive orthant of a sphere, where
\begin{equation} \mathbf{q} = (q_1,\ldots,q_{d+1})^T = \left( \frac{x_1^\alpha}{\| \mathbf{x}^\alpha \|} , \ldots, \frac{x_{d+1}^\alpha}{\| \mathbf{x}^\alpha \|} \right)^T \; \in \; S_+^d, \label{powertrans}
\end{equation}
where $\| \mathbf{x}^\alpha \| = \sqrt{ \sum_{j=1}^{d+1} x_j^{2\alpha} } $ is the Euclidean norm of $\mathbf{x}^\alpha$, and $S_+^d$ is the positive orthant of the sphere, where each $q_j \ge 0$ and $\sum_{j=1}^{d+1} q_j^2 = 1$. We can choose the power $\alpha$ to carry out our analysis, for example $\alpha=\frac{1}{2}$ gives the square root transformation to the sphere of \citet{Scealwels11} and in this case 
$ \| \mathbf{x}^{1/2} \| = 1$.  The value $\alpha=1$ is used by \citet{Lietal23}. 
After transformation we can apply methods for the analysis of spherical data analysis \citep{Mardjupp00} to analyze compositional data, and then we map back the results from the positive orthant of the sphere to the simplex. The
inverse transformation is given by 
$ x_j = {q_j}^{1/\alpha} /{ \sum_{j=1}^{d+1} q_j^{1/\alpha}   } , \; j=1,\ldots,d+1. $

Throughout, we use the square-root embedding with $\alpha=1/2$
since Euclidean geometry on the sphere then corresponds to the Fisher–Rao metric on the simplex which is the canonical Riemannian metric for probability distributions. Under the square-root transformation $q_j=\sqrt{x_j}$, the Fisher-Rao metric 
$ds^2=\sum_{j=1}^{d+1} \frac{dx_j^2}{x_j}$ reduces to $ds^2=4\sum_{j=1}^{d+1} dq_j^2$, 
so that the simplex with the Fisher-Rao geometry is isometric (up to scale) 
to the positive orthant of the unit sphere with the Euclidean metric \citep{Srivastavaetal07,Srivklas16}. This 
choice of induced metric is invariant under permutation of components, is symmetric in all coordinates,
and independent of arbitrary scaling choices, and so it is a natural choice of distance. In practice we can use either geometric justifications or data-based choices for $\alpha$, where $\alpha = 1/2$ often works well as a variance stabilizing choice.

Note that there are $d+1$ components in the composition, which leads to a simplex of dimension $d$ ($\Delta^d$) and a sphere of dimension $d$ ($S^d_+$), since the constraints $\sum_{j=1}^{d+1} x_j = 1$ and 
$\sum_{j=1}^{d+1} q_j^2 = 1$ each impose a single linear or quadratic constraint, respectively.

\subsection{Principal Nested Spheres}\label{PNSsection}
We focus on the dimension reduction technique of Principal Nested Spheres (PNS) \citep{Jungetal12}, and a recent detailed treatment of high-dimensional PNS has been given 
by \citet{Monemetal25}. 
The method of PNS is a non-linear dimension reduction technique for spherical data. The method is a backwards fitting procedure, starting with fitting a high-dimensional subsphere to a spherical dataset and then fitting further subspheres, successively reducing dimension at each stage. 
Let us write $\Phi$ for the set of PNS parameters \citep{Monemetal25}, where 
\begin{equation} {\Phi} = \{ v_1,\ldots,v_d,r_1,\ldots,r_{d-1} \} . \label{pars}
\end{equation}
where $v_k \in S^{d+1-k}$ when starting from data on $S^d$ are axis directions that are orthogonal to the plane that defines the subsphere, and 
$0 < r_k \le \pi/2$ are angles between the subsphere and the axis. At each stage of the fitting procedure we choose between $r_k = \pi/2$ for a great subsphere and $r_k < \pi/2$ for a small subsphere. Least squares fitting is commonly used for estimating the parameters, and methods for choosing between a great subsphere and small subsphere at each stage include the Bayesian Information Criterion (BIC), a sequential Bootstrap test, a Kolmogorov-Smirnov (KS) test, a variance test, and a likelihood ratio test \citep{Jungetal12,Monemetal25}.

If we have a sample size $n$ of observations $x_i^{(d)}, i=1,\ldots,n$  on $S^d$ (where $n>d$), we would like to compute the PNS scores given the PNS parameters, and these are obtained from the residuals at each stage of the fitting \citep{Jungetal12}. In total there are $d$ PNS scores for calculating PNS on a sample on $S^d$, where $d$ is also the dimension of the sphere. 

We denote the PNS scores for the $i$th observation as the $d$-vector:
\begin{equation}
(s_{1,i}, s_{2,i}, \ldots, s_{d-1,i}, s_{d,i})^T, \; i=1,\ldots,n.  \label{PNSscores}
\end{equation}
The first score is a circular variable and the other scores are defined on intervals. The radius of the circle for PNS score 1 and the widths of the intervals are determined from the PNS parameters $\Phi$. So, the space of the PNS scores is in a PNS-cylinder space ${\cal C}^d$ \citep{Monemetal25}. 
This space is a product of a circle (for the first score) and Euclidean intervals for the remaining scores.

Note that it is the positive orthant that is of primary interest, which can then be mapped back to the simplex.
However, familiar methods for spherical data analysis are not restricted to the positive orthant, for example a density function on the sphere or fitted subspheres from PNS. So one must address this issue 
before transforming back to the simplex, and a convenient approach is to use truncation as in the next example.

\subsection{Application: 3D compositional data}\label{application}
To demonstrate the application to compositional data analysis we consider some 
geochemical data in 87 samples of rocks in Western Australia, which is a subset of the data set of \citet{Rcomp}. We consider a composition of three variables: Chromium (Cr), Zinc (Zn), and Lead (Pb), and we normalize them to proportions that add to 1 on the simplex  $\Delta^2$. 

\begin{figure}[htbp]
     \centering
     \includegraphics[width=7cm]{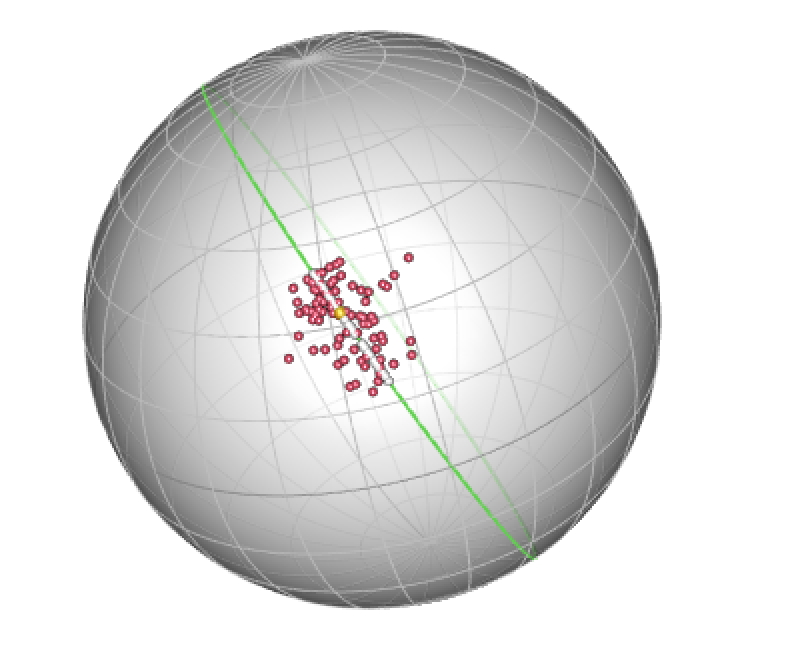}
    \includegraphics[width=7.2cm]{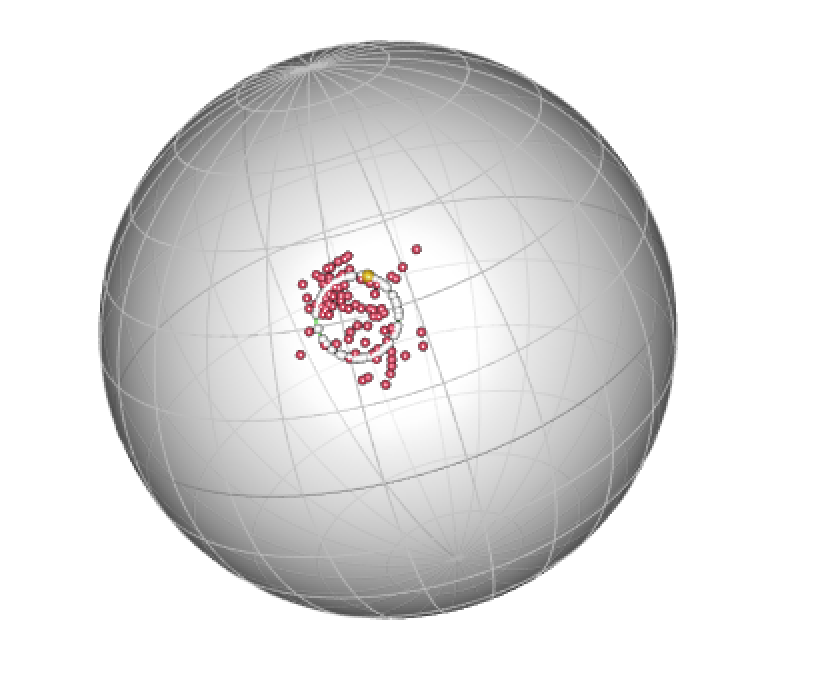}
\caption{The great circle fit (left, in green) and the small circle fit (right) for the 3D geochemical data using power $\alpha=1/2$. 
The green solid line is the fitted subsphere and each point is projected onto the subsphere (in white). 
Also, shown in yellow is the PNS mean.}\label{great-small}
\end{figure}

\begin{figure}[htbp]
\hskip -0.5cm  \includegraphics[width=\textwidth]{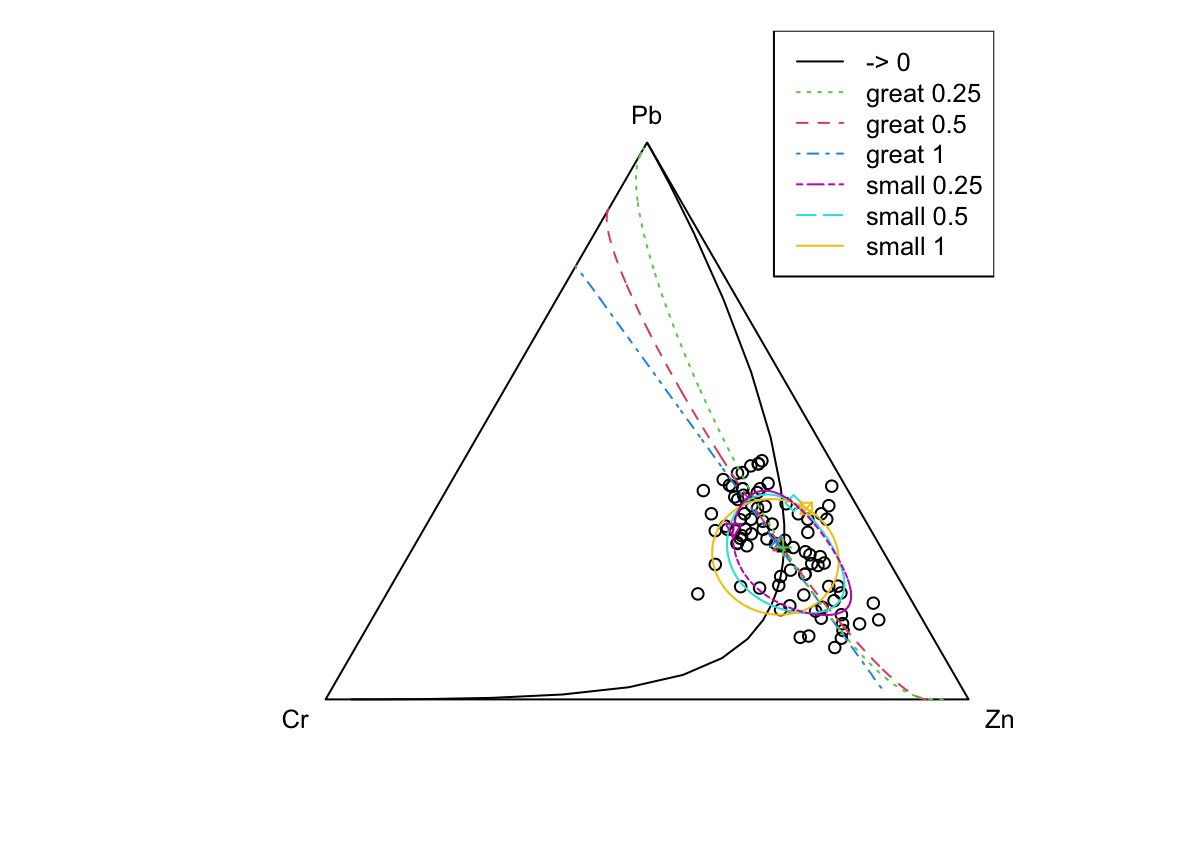}
     \caption{A ternary diagram with fitted PNS subspheres. The first principal component from the {\tt compositions} package is shown in a black solid line, and corresponds to $\alpha \to 0$. 
     The great circle fits for $\alpha \in \{ 0.25, 0.5, 1\}$ are given by green, red and blue dashed lines, and the small circle fits 
     for $\alpha \in \{ 0.25, 0.5, 1\}$ are given by purple, cyan and gold lines. The great and small circle PNS means are given by the respective coloured symbols.}
     \label{ternary}
\end{figure}

We are interested in describing the mean and variability of the data. 
First of all, we take the power transformation of the data from (\ref{powertrans}) to represent the data on $S^2$, 
and focus on $\alpha \in \{ 0.25, 0.5, 1 \}$. 
We carry out PNS on this dataset using a great subsphere (a great circle) and a small subsphere (a small circle) using the command {\tt pns} in the {\tt shapes} library in R \citep{Rshapes,Drydmard16}. 
For $\alpha=1/2$ in Figure \ref{great-small} we  plot the data (red circle points) on the sphere $S^2$ with a great circle fit (left-hand plot) and a small circle fit (right-hand plot). 
Each fitted subsphere is then truncated to $S^2_+$ and then projected back to the simplex in a ternary diagram in Figure \ref{ternary}, together with the results for powers $\alpha=0.25$ and $\alpha=1$.  

%The first PNS score for the great circle explains approximately $67.7\%$ of the variability and the second PNS score explains $32.30\%$. The first PNS score for the small circle explains approximately $93.92\%$ of the variability and the second PNS score explains $6.08\%$.

In the ternary diagram the coordinates
in the triangle are given by the perpendicular distance to the edge opposite 
each corner. For example, a point in the center would have coordinates for (Cr, Zn, Pb) equal to  $(1/3,1/3,1/3)$, a point 
halfway along the bottom edge would have co-ordinates $(1/2,1/2,0)$ and a point at
the Zn corner would have co-ordinates $(0,1,0)$.
In Figure \ref{ternary} we see the data 
points plotted in the ternary diagram for the
87 observations. For all observations, the Zn component is the largest, as the data points are all furthest from the Cr-Pb edge. The PNS mean composition using the great circle fit with $\alpha=1/2$ is at $(0.153,0.571,0.275)$, showing a mean composition of mostly Zinc, then Lead, and then smallest proportion of Chromium. 

Visually the arcs on the simplex corresponding to the fitted great subspheres are quite similar through the range of the data for $\alpha \in \{ 0.25, 0.5, 1 \}$. The $\alpha =1$ great circle has the advantage that it projects to a straight line 
on the simplex, which can aid interpretation. 
The fitted small subspheres are also quite similar to each other.

A traditional choice of principal components analysis (PCA) for compositional data is computed from the centered log-ratio transform (clr) in the R package {\tt compositions} \citep{Rcomp},  and the PC1 is plotted on the ternary diagram as the black curve. Although the curve passes through the centre of the data the PC does not really seem to capture the structure of the variability well, and so the great subsphere summaries look preferable.

\section{Manifold Regression}\label{comp reg}
\subsection{Intermediate spaces}
Regression analysis is one of the most common statistical tools for modelling relationships between variables and predicting future values. In a regression model, there is a set of predictors (independent variables) and a set of responses (dependent variables), most often assumed to have a linear relationship. For more general relationships a link function is used to model the relationship between the two sets of variables, which
is useful for parameter estimation and prediction of future values. In our study, we want to develop methods for regression analysis for complex data objects that lie on nonlinear manifolds. This methodology involves mapping the complex data into { intermediate spaces} where a linear regression model can be formed between the transformed predictors and responses.

Consider a data set of paired observations $\{(\mathbf{X}_i, \mathbf{Y}_i), i=1,\ldots,n\}$ where $\mathbf{X}_i \in \mathcal{M}_X$, the predictor manifold,
and $\mathbf{Y}_i \in \mathcal{M}_Y$, the response manifold. The aim is to estimate a regression function $f: \mathcal{M}_X \rightarrow \mathcal{M}_Y$ either parametrically or nonparametrically.

We define a map of the response variable from ${\mathcal M}_Y$ to an {intermediate space} 
 ${\cal L}_Y$ and 
suitable transformation from ${\mathcal M}_X$ to  another intermediate space ${\cal L}_X$. We make this structure explicit by introducing maps
\[
h : \mathcal{M}_X \rightarrow \mathcal{L}_X, \qquad
g : \mathcal{M}_Y \rightarrow \mathcal{L}_Y,
\]
where $\mathcal{L}_X$ and $\mathcal{L}_Y$ are statistically tractable intermediate spaces, and $g$ is bijective. The inverse map is $g^{-1} : \mathcal{L}_Y \rightarrow \mathcal{M}_Y$. Regression is then defined through
\[
f(\mathbf{X}) = g^{-1}\!\bigl(w(h(\mathbf{X}))\bigr),
\]
where $w : \mathcal{L}_X \rightarrow \mathcal{L}_Y$ is a regression function in the intermediate spaces. Note that $g^{-1}$ plays the analogous role of a link function that is used in classical regression. 

\subsection{Simplex on Euclidean regression using PNS intermediate space}
Our study will focus on the case where the response manifold is a simplex for compositional data ${\cal M}_Y = \Delta^d$ and the predictor space is Euclidean ${\cal M}_X = {\mathbb R}^p$.
Let $\mathbf{\tilde Y} = g( \mathbf{Y} )$ be a $k$-vector in ${\cal L}_Y$ and let $\mathbf{\tilde X} = h(\mathbf{X})$ be a $m$-vector in ${\cal L}_X = \mathbb{R}^m$. 

In the PNS-based regression considered here
the function $g: {\cal M}_Y \rightarrow {\cal L}_Y$ is obtained by first mapping to the positive orthant of the sphere using (\ref{powertrans}), then carrying out PNS decomposition to the PNS scores. Hence ${\cal L}_Y$ is a PNS cylinder space ${\cal C}^d$ of $k=d$ PNS scores (\ref{PNSscores}), with the PNS parameters $\Phi$ given by (\ref{pars}). Full details of this space are given by 
\citet{Monemetal25}.

We consider the predictor manifold ${\cal M}_X$ to be Euclidean with $p$ real-valued predictors and an intercept, and so $h(\mathbf{X})= \mathbf{\tilde X} = (1,\mathbf{X}^T)^T$ here, which is a $m$-vector with $m=p+1$. Hence, the intermediate 
regression function $w$ needs to be estimated between the predictor space $\mathbb{R}^m$ and the PNS-cylinder space ${\cal C}^d$, described after (\ref{PNSscores}).

The full regression model involves estimation 
of the PNS parameters $\Phi$ from (\ref{pars}) in order to specify 
the function $g$, which we denote $g_{\Phi}$. 
Our simplex-Euclidean model is
therefore:
$$ \mathbf{Y} = g_{\Phi}^{-1} ( \mathbf{B} \mathbf{\tilde X} + \mathbf{E} ) \; \; , $$
where $\mathbf{B}$ is a $k \times m$ matrix of regression parameters and $\mathbf{E}$ is an  zero-mean random $k$-vector. 

We will consider a two-stage procedure to fit the model to an observed dataset 
of paired observations $(\mathbf{X}_i, \mathbf{Y}_i), i=1,\ldots,n$. First, we fit the PNS 
decomposition parameters $\hat\Phi$ using the response data $\mathbf{Y}_1,\ldots, \mathbf{Y}_n$ as an initial part of the estimation procedure. 
The second stage of the model fitting is then carried out conditional on this PNS fit, to obtain the estimated linear regression parameters $\mathbf{\hat B}$.
After fitting the intermediate space model with
fitted regression response $\mathbf{\hat w} \in {\cal L}_Y$ the mapped back response on ${\cal M}_Y$ is then $\mathbf{Y} = g_{\hat\Phi}^{-1}( \mathbf{\hat w}( h(\mathbf{X}) ) = g_{\hat\Phi}^{-1} ( \mathbf{B} \mathbf{\tilde X} )$ here.

The fitting procedure at the second stage is not entirely straightforward as the first PNS score is a circular variable, and so the wrapping of 
the first PNS score onto a circle needs to 
be considered at all stages of the analysis. 
Here we have described using all $d$ PNS scores. However, in practice, the number of PNS scores may be selected using explained variability or prediction performance via cross-validation.

\subsection{Simulation study}
We carry out a simulation study to examine some methods for regression where the response is a simplex and we have Euclidean predictors. The specific simulation model is designed as follows. We consider the geochemical data of \citet{Boogtolo13}
with particular composition (Cu,Ga,Nb,Ni,Pb) transformed to the positive orthant of the sphere using the square root function. We carry out PNS using small subspheres to give the fitted PNS parameters $\Phi^*$, on which the simulation is based.  A sample of $n=100$ observations is then obtained by simulating PNS scores:
\begin{eqnarray}
s_{1,i} & = & 0.1 + 1.6 x_{1i} + 0.4 x_{2i} + \epsilon_{1,i}\\
s_{k,i} & = & \epsilon_{k,i} , \; \; \; k=2,\ldots,d
\end{eqnarray}
where $x_{1i} = (-51+i)/100, x_{2i} = \sin( 2 \pi (-51+i)/100) $ for $i=1,\ldots,n$ with additive errors $\epsilon_{l,i} \sim {\rm iid} N(0,\sigma^2), l=1,\ldots,d$ with $\sigma=0.05$. 
These PNS scores are then back-transformed to the orthant of the sphere, and squared to give the simulated compositions $\mathbf{z}_i, i=1,\ldots,n$. 

We then fit various regression models. 
In brief the methods are: 

\begin{enumerate} 
\item PNS score 1 circular-linear regression. Transform the composition to the sphere, carry out PNS giving the intermediate space as the space of PNS scores. So 
$g$ here is the map from composition simplex to the PNS score space. Then we carry out circular-linear regression on PNS score 1 using the known predictors. 
One of the issues is that PNS score 1 $s_{1i}$ is a circular variable, with range  $[-r \pi, r \pi)$ where $r$ is the radius of the 
PNS 1 circle and $i=1,\ldots,n$. 
So, with some larger regression coefficients there can 
be winding round the circle past the cut point. Thus what might be a linear 
function has a big jump in it when the angle jumps by $\pm 2\pi r$ radians. 
In order to address this problem we need to take into account the circular nature of the response for PNS score 1. Let us write $y_i = s_{1i}/r \in [-\pi,\pi)$ for the circular variable.
In particular, we consider the
model
$$ y_i + \pi = (m_i + e_i + \pi) {\rm mod} 2\pi , $$
where $m_i = \mathbf{\tilde X}_i^T \beta$ is a circular mean dependent on predictors and $\beta$ is a $p+1$-vector of regression parameters. This formulation ensures continuity across the cut point and allows least-squares estimation while respecting the circular nature of the response.
Note $y_i \in [-\pi, \pi)$. For the leading PNS score 1 circular variable we use a least squares fit. 
The fitted values for PNS score 1, together with zeros for the other non-leading PNS scores, are then back-projected to the sphere and then squared to the composition simplex. 
This transformation from PNS scores to the simplex is the inverse map denoted by $g^{-1}$. 

\item PNS all scores circular-linear regression.
The method is exactly the same as in the first method except we use all the PNS scores for the back projection, not just the first. 
\item PNS circular package c-l. The method is similar to the first approach except the circular-linear method of the {\tt circular} package is used for the circular-linear regression. The model assumes that a circular response variable  has a von Mises distribution with concentration parameter $\kappa$, and mean direction related to a vector of linear predictor variables.
\item Linear regression on simplex. For this method we just use linear regression on each component of the simplex separately. 
\item Quadratic regression on simplex. Here quadratic regression is used on each separate component. 
%\item PNS to simplex linear regression. For this method the PNS score 1 is transformed back to the simplex and then separate linear regressions are performed. 
\item PCA score 1 linear regression. PCA is carried out on the compositional data treated as Euclidean vectors (a naive baseline approach), regression is carried out on PC score 1, and then mapped back to the composition.
\item PCA score after arcsine transformation. The arcsine transformation is carried out on the compositions, PCA carried out, then regression performed on PC score 1 and mapped back to the composition. 
\end{enumerate}

\begin{figure}[htbp]
\begin{center}
\includegraphics[width=16cm]{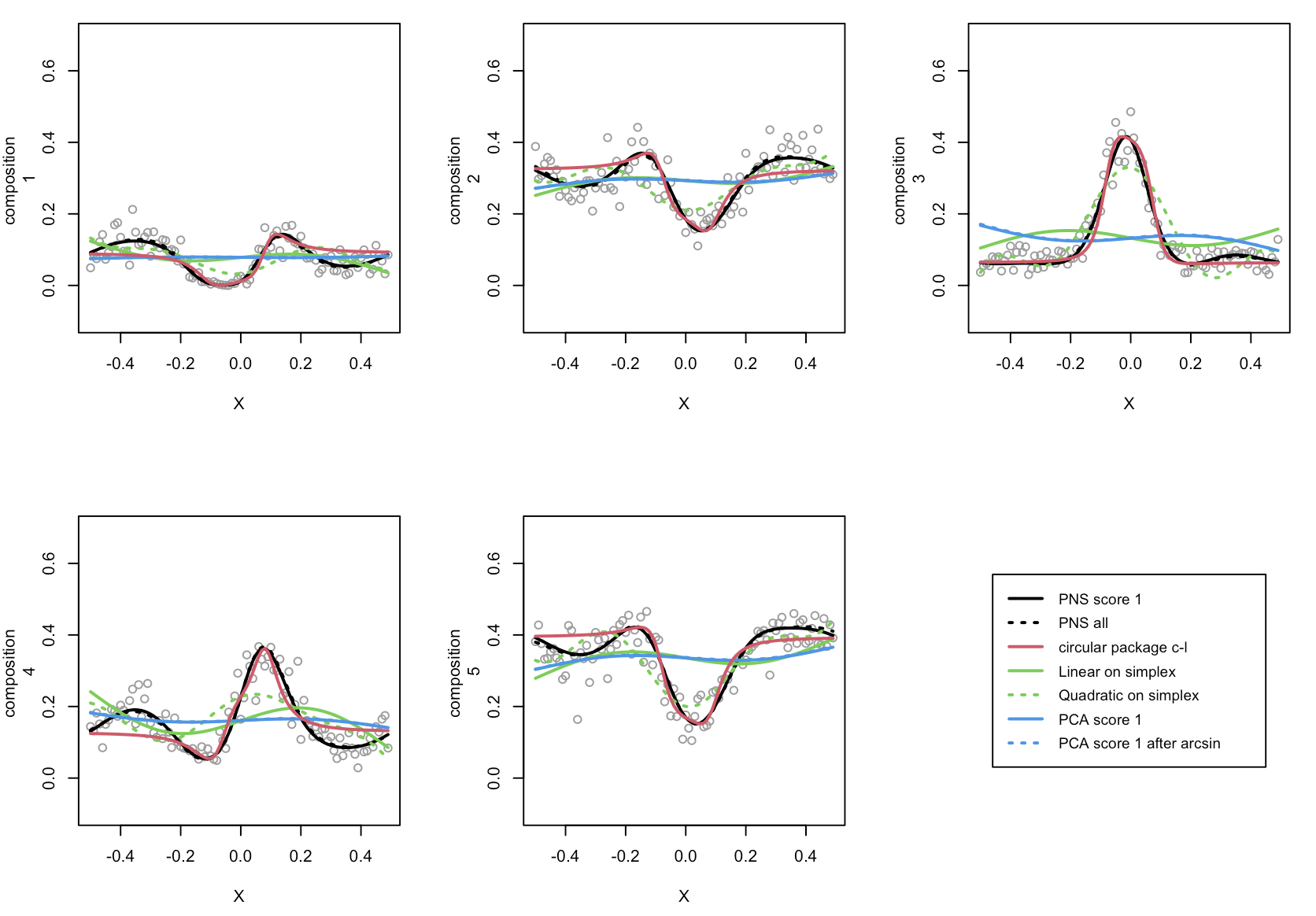}
\end{center}
\caption{Simulated compositional data with five components. The `PNS score 1' and `PNS all' methods provide particularly close fits, with the circular package the next best, and the remaining methods are all rather poor. }\label{simulation-results}
\end{figure}

Each of the methods is fitted to the simulated data and a plot of the fitted values for the composition versus 
the first predictor are given in Figure \ref{simulation-results}. We see that the PNS methods both do very well in fitting the data. The circular package method does not capture the bimodal effect as well, and all the linear methods are not able to fit the peaks very well.

\begin{table}[htbp]
\centering
\begin{tabular}{ |p{8cm}||p{3cm}|}
 \hline
 \multicolumn{2}{|c|}{PMSE after cross validation} \\
 \hline
 Different methods & PMSE\\
 \hline
PNS score 1 circular-linear regression &   {\bf 14.85} \\
PNS all scores circular-linear regression &   15.00  \\
PNS circular package c-l & 24.21 \\
Linear regression on simplex & 66.12 \\
Quadratic regression on simplex  & 34.01 \\
%PNS to simplex linear regression &  66.01\\
PCA score 1 linear regression& 69.09 \\
PCA score 1 after arcsine  linear regression &  69.13  \\

\hline
\end{tabular}
\caption{A table showing results of cross-validation of the different regression methods, with random splits of 80\% of the data used for fitting, and 20\% used for prediction, with prediction mean square error (PMSE) reported.}
\label{TAB1}
\end{table}

We carry out cross-validation using 80\% training and 20\% test data, and the results are given in Table \ref{TAB1}. Clearly here the PNS method with least squares circular regression performs the best. 
Although the data are generated from a PNS-based simulation, the comparison remains informative because competing methods fail even to recover the dominant nonlinear structure.

\section{Application: Environmental contaminants data}\label{EPA_project}
\subsection{Exploratory data analysis}
The proposed PNS regression methodology was applied to a dataset that was compiled from an Environmental Protection Agency (EPA) funded project \citep{Ogunbiyietal24}. The study examined contaminant chemicals that were detected in soil, food, water, dust, and urine samples from children in South Florida, USA. The study consists of 57 children with two years follow-up by quarters which means we have up to eight data collections from the same children with 145 variables, representing various chemical peak areas, demographic information, and survey data. The dataset contains chemical peak area measurements from mass spectrometry categorized into five primary sources which are water (11 chemicals), dust (27 chemicals), food (13 chemicals), soil (7 chemicals), and urine (3 chemicals). The number of chemicals is chosen based on chemicals found in more than $50\%$ of samples, except for urine where the three chemicals are chosen as specific chemical tracers in soil and dust ingestion \citep{Ogunbiyietal24}. 
We can consider that water, dust, food and soil are input sources of chemicals, and urine chemicals are potential output sources. 

\begin{figure}[htbp]
\begin{center}
\includegraphics[width=14cm]{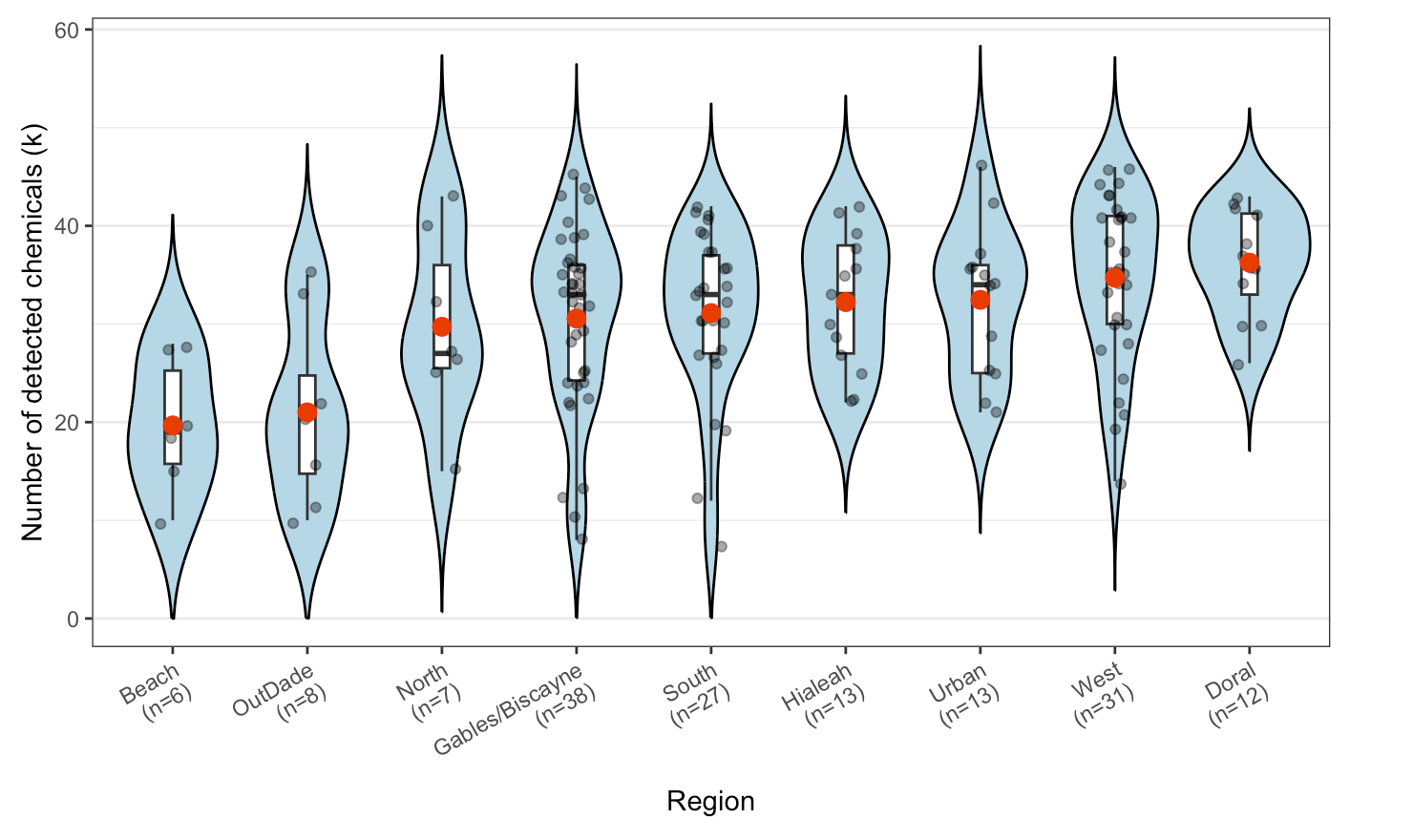}
\end{center}
\caption{The number of chemicals detected in total over  the matrices they represent (water, dust, food and soil) in the nine grouped geographical areas in the study. The distribution of the number 
of detected chemicals is displayed as a violin plot for each of nine areas with the mean in red, an underlying boxplot and jittered observed values.}\label{violin}
\end{figure}

\begin{figure}[htbp]
\begin{center}
\includegraphics[width=6cm]{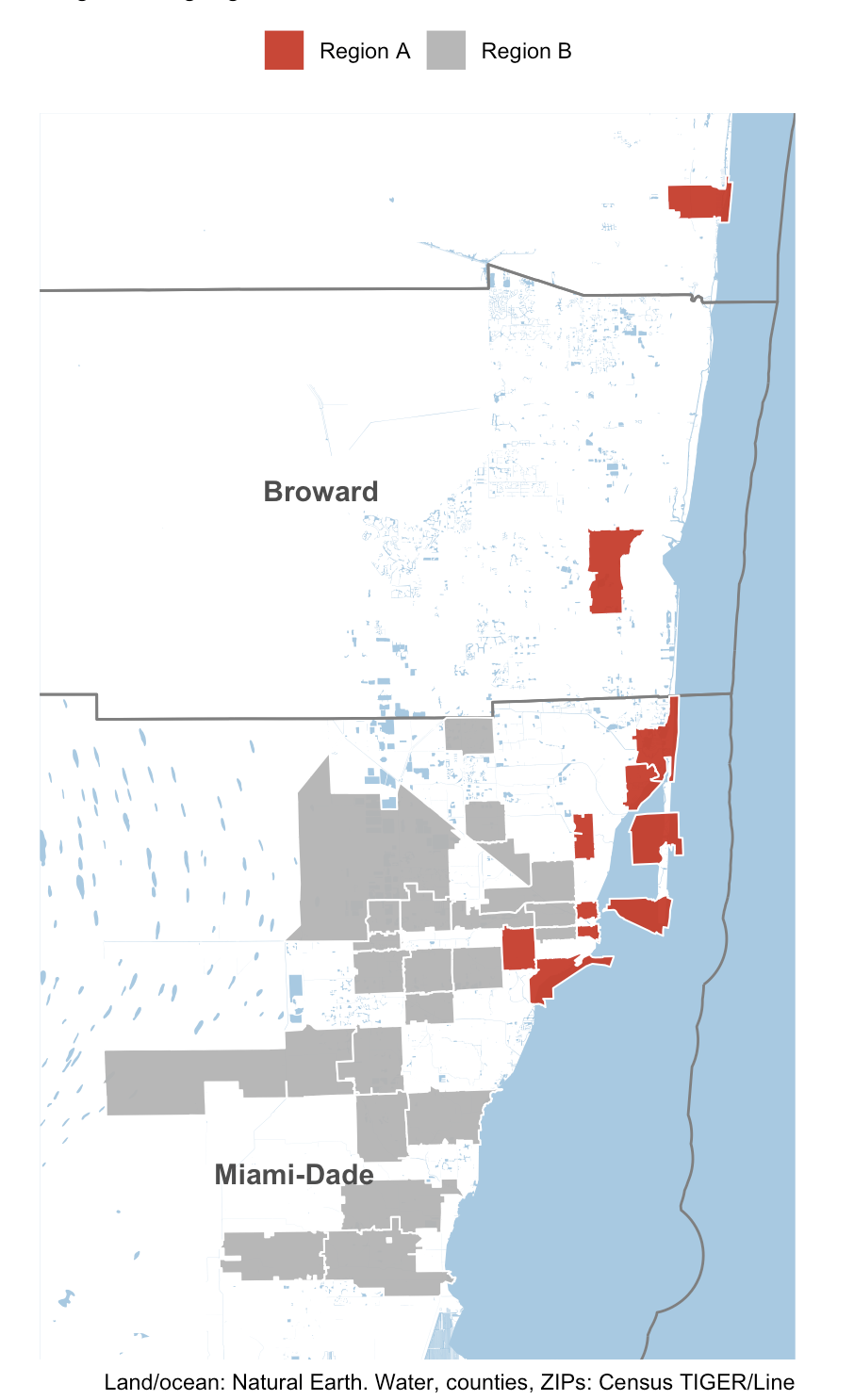}
\end{center}
\caption{A map of the zip codes in Region A and Region B. Region A locations are mainly along the coast or near to Biscayne Bay}\label{map}
\end{figure}

The obtained chromatographic peak areas representative of each identified chemical (detected by mass spectrometry based on accurate mass to charge ratio) were compared between children for each chemical. The comparison between different chemicals is challenging as they can exhibit distinct ionization efficiencies, resulting in different responses in the mass spectrometer under identical conditions. So, in order to carry out meaningful compositional data analysis we first convert the peak area to a binary measure: 1 if the chemical is detected (peak area greater than zero) and 0 if not. Then we calculate a single composition vector for all of the water, dust, food, and soil chemicals so that the sum of the measures over all the chemicals is 1. Hence, the composition is a measure of chemical diversity: larger diversity would give fewer zero components. The number of detected chemicals is given in Figure \ref{violin} for different locations in the study. Zip codes are grouped into nine  geographical areas. Four geographical areas have lower mean counts of chemicals per collection than others: the more coastal areas Beach, OutDade, North, Gables/Biscayne which we group into Region A, and the other areas are grouped into Region B (mainland Miami-Dade). 
A map of the zip codes is shown in Figure \ref{map}.

\begin{figure}[htbp]
\begin{center}
\includegraphics[width=15cm]{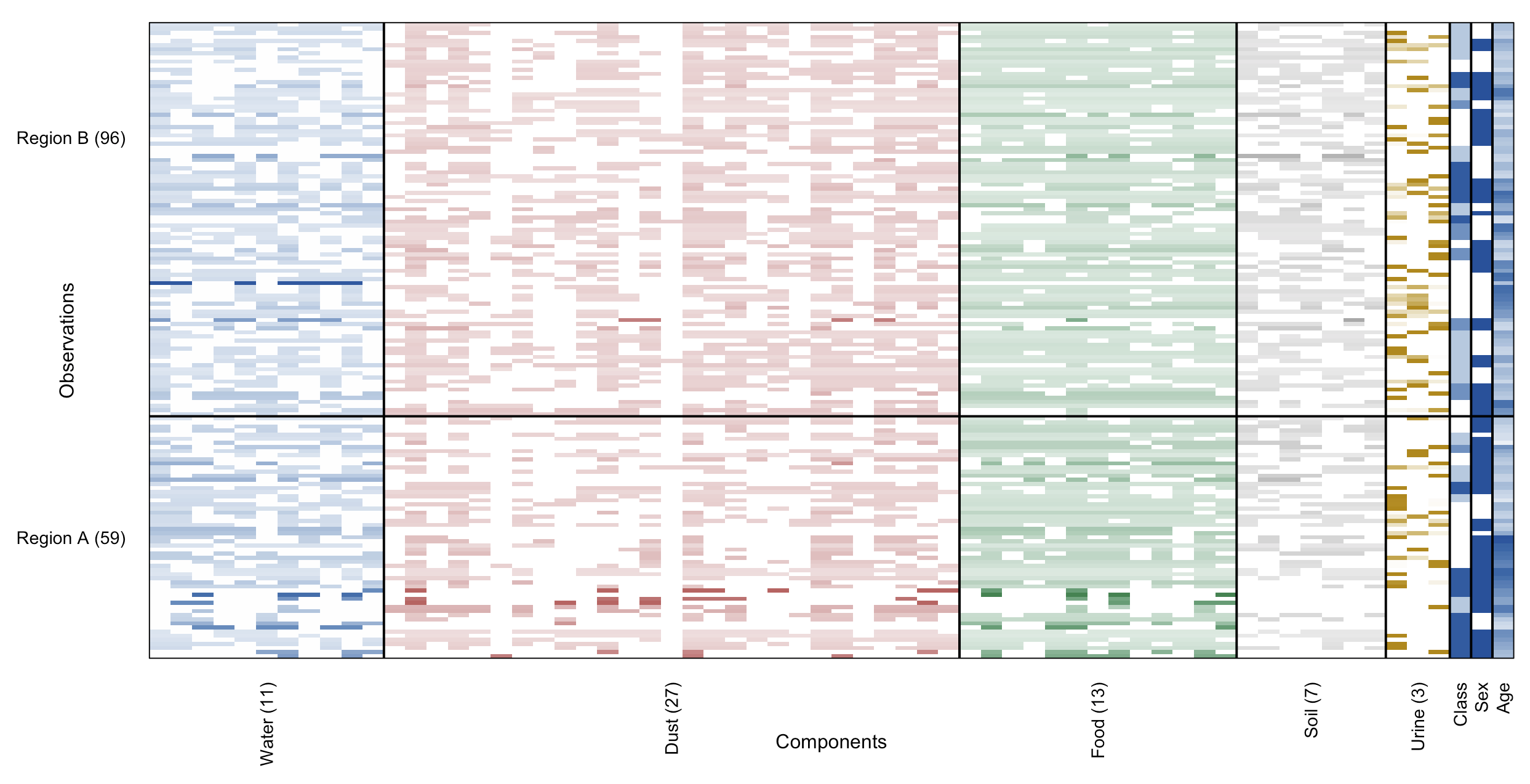}
\end{center}
\caption{The environmental chemical data with each row of the water, dust, food and soil vector mapped to the positive orthant of the sphere $S^{d}_+$, where $d=58$. The indicators of chemical presence for water, dust, food, and soil are indicated with darker colours for higher composition values. The observations are in the rows, ordered by location (Region A below and Region B above). The number of chemicals for each data source are given in parentheses below the columns. The urine composition is also included for columns 59, 60, 61 and final three columns are the economic class (white for Low, light shade for Middle and NA and dark shade for Upper), and for the urine data: sex (white for Boy and dark shade for Girl) and age (darker for older ages).}\label{envdata}
\end{figure}

We then transform the compositional data to the positive orthant of the sphere using $\alpha=1/2$ in Equation (\ref{powertrans}). After removing incomplete data there are $n=155$ observations with all $d+1=58$ input chemicals tested with 59 measurements from Region A and 96 from Region B. Compositions for the three urine chemicals are also computed (although 23 urine samples were not available). There is also information about the economic class (Low, Middle, Upper and Not Available (NA)), and the sex and age (in months) for the children giving urine samples (51 measurements from boys, 81 measurements from girls) with ages from 13 to 83 months. The dataset is displayed in Figure \ref{envdata}. 

\subsection{PNS regression}
\begin{figure}[htbp]
\begin{center}
\includegraphics[width=14cm]{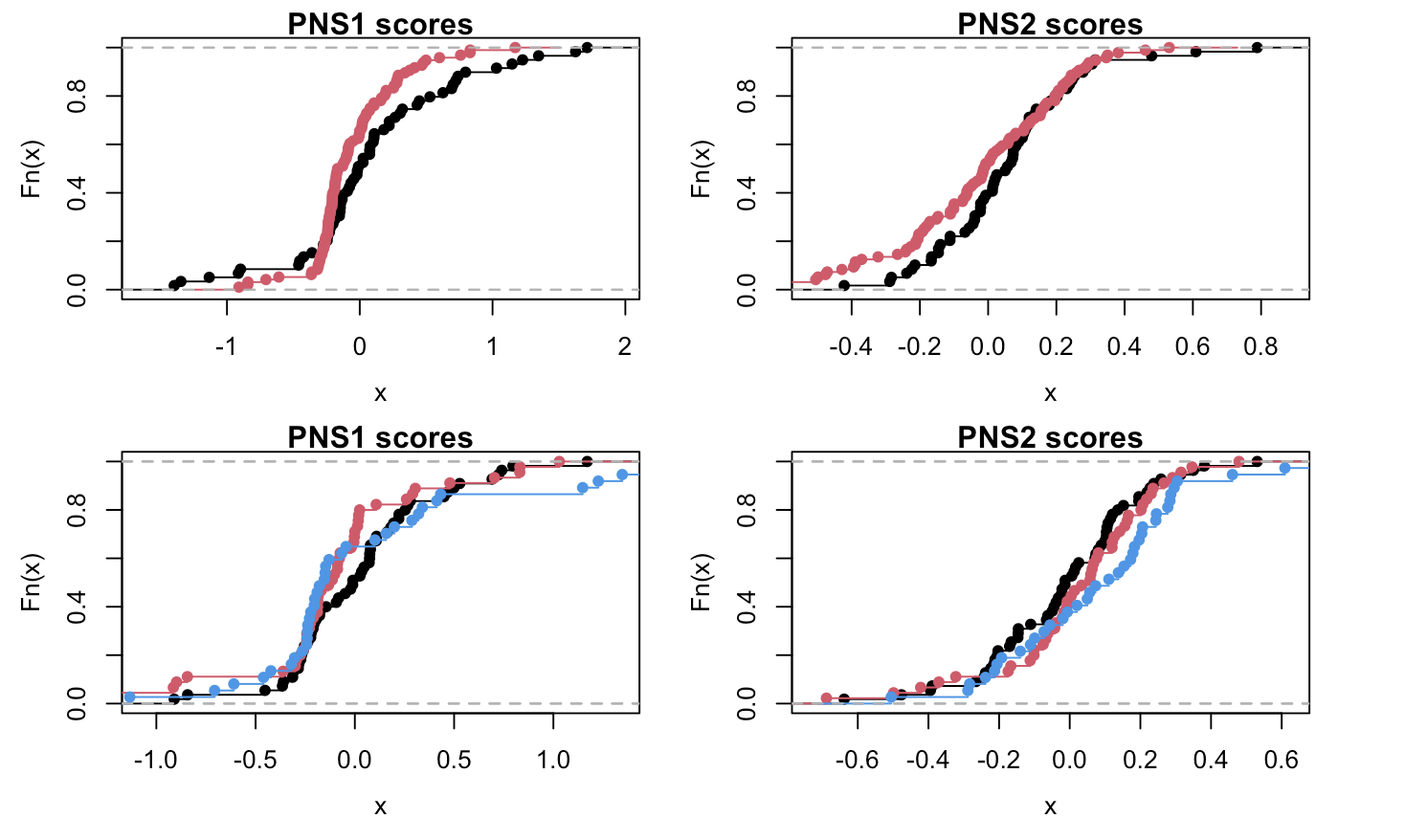}
\end{center}
\caption{Above: the empirical cdf of the first two PNS scores for Region A (black) and Region B (red), clearly showing that Region A have higher PNS1 scores on average and higher PNS2 scores on average. Note that Region A has a larger variance in PNS1 than Region B. 
Below: the empirical cdf of the first two PNS scores by economic class: Low (black), Middle (red), Upper (blue), showing that PNS2 score is higher for the Upper class than the Low class on average. 
}\label{ecdf}
\end{figure}

We apply PNS to the transformed data on the sphere using the Bayesian Information Criterion to select the angles $r_k, \; k=1,\ldots,d-1$ \citep{Jungetal12,Rshapes}. The first three PNS scores explain 33.2\%, 8.5\% and 5.7\% of the variability, respectively. 
Empirical cumulative distribution functions (cdfs) for the PNS1 and PNS2 scores are given in Figure \ref{ecdf}, which illustrate the differences between the groups for PNS1 score (Region A higher than Region B on average) and PNS2 score (Region A higher than Region B on average; Upper class higher than Low class on average).

We fit a circular-linear regression model for PNS score 1 versus location (Region A, Region B) economic class (Low, Middle, Upper, NA) and we also include the sex (Boy, Girl) for the child. 
There is some evidence that the region has a significant effect on PNS1 (Region A higher than Region B: p-value = 0.03), but there is not a significant class or sex effect.  
For the linear regression model of PNS score 2 there is weak evidence for a location effect (Region A higher than Region B: p-value = 0.1), and 
a class effect (Upper Class higher than Low class: p-value =0.08), but no sex effect.

Given that we have longitudinal data with up to eight quarterly measurements per child/household in the study, we also fit mixed effect models using {\tt lmer} in the {\tt lme4} library in R \citep{Bates05}. The model includes a random intercept and an uncorrelated random slope for quarter at the child/household level, allowing both the starting level and the rate of change over time to vary between children/households. We then test whether the fixed effects are significant using the Kenward-Roger method \citep{Kenwroge97}. There is some weak evidence that the region has a significant effect on PNS1 (Region A higher than Region B: p-value = 0.064), but no class or sex effect.  
For PNS score 2 there is a class effect (Upper class higher than Low class: p-value =0.014), but no area or sex effect.  

So, from both analyses the common aspects are that PNS1 score is significantly higher for Region A than Region B, and PNS2 score 2 is significantly higher for Upper class versus Low class.

\begin{figure}[htbp]
\begin{center}
 \includegraphics[width=15cm]{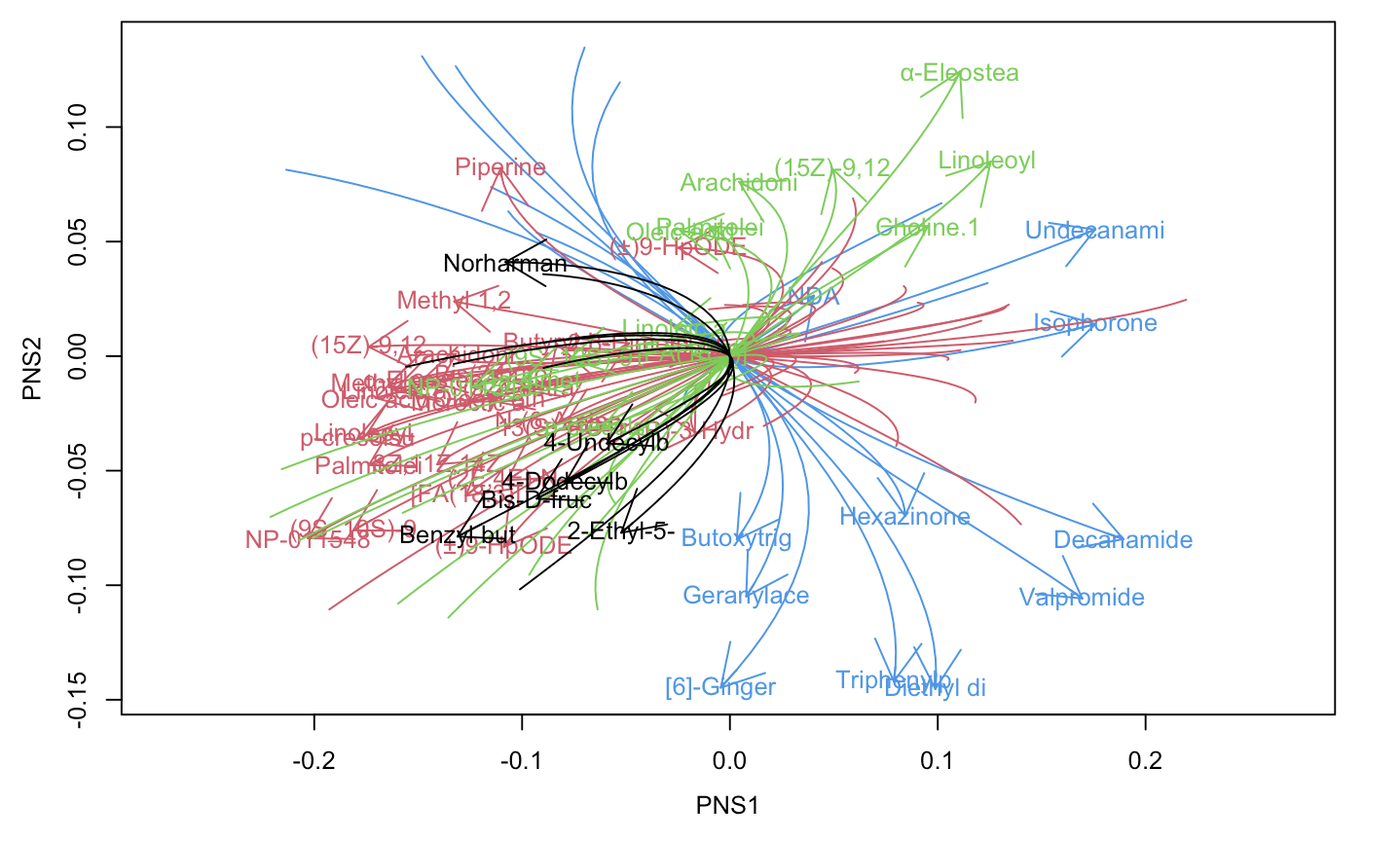}
\includegraphics[width=14.9cm]{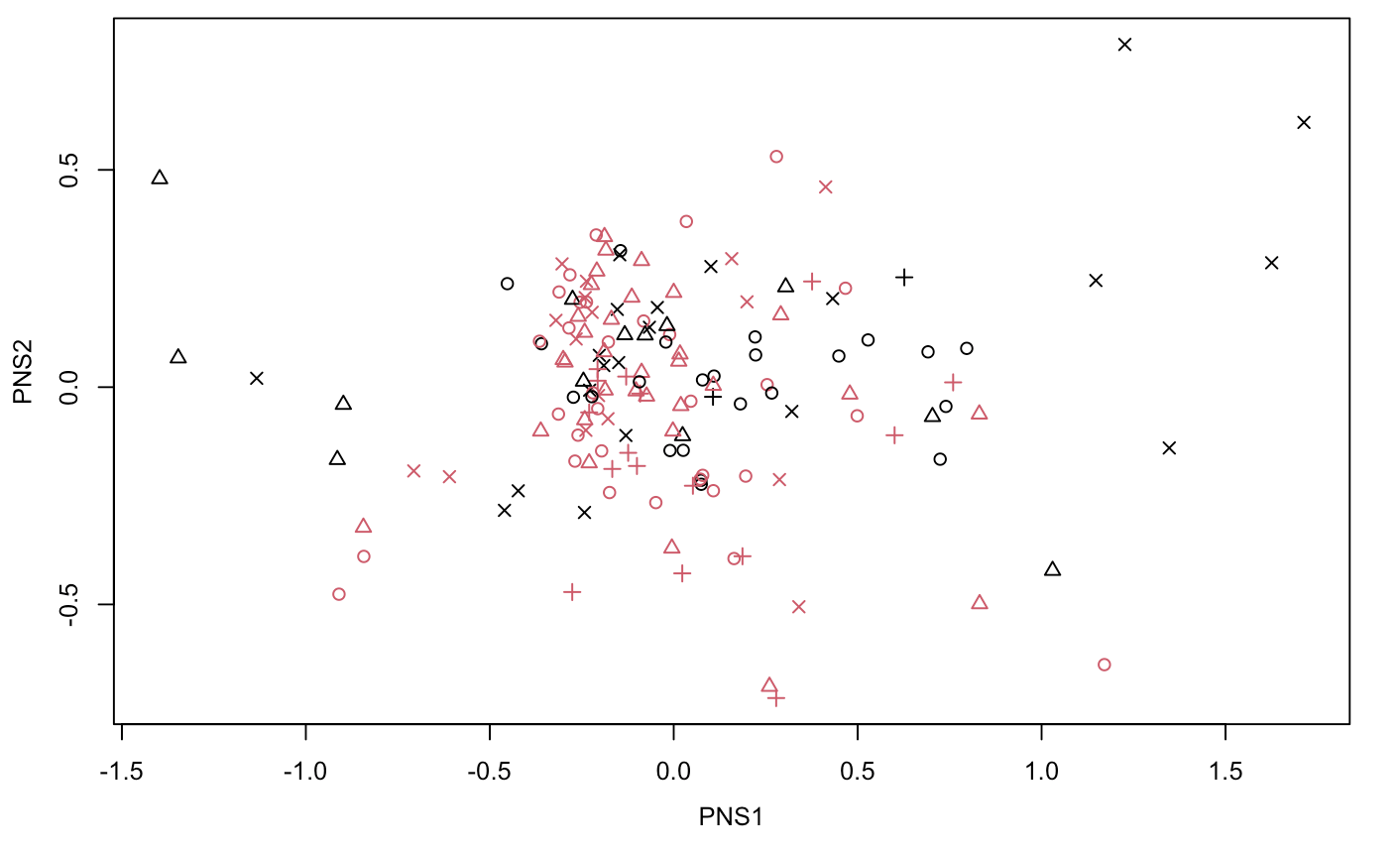}
\end{center}
\caption{The PNS biplot for the environmental chemical data. (Above) Paths showing the relative contributions of the chemicals in the PNS1 and PNS2 directions:  
water (blue), dust (red), food (green), soil (black).  (Below) The first two PNS scores labeled by location (Region A black, Region B red) and by economic class ($\circ$ Low, $\triangle$ Middle, $\times$ Upper, $+$ NA). }\label{PNS2}
\end{figure}

The PNS biplot is a graphical method to assess which of the original variables contributes most to the PNS scores \citep{Monemetal25}, and it has a similar role to the traditional biplot \citep{Gabriel1971Biplots}. 
From the upper plot in Figure \ref{PNS2} we see that PNS1 shows a contrast between dust chemicals (lower PNS1, higher number of dust chemicals detected) and food/water chemicals (higher PNS1, higher numbers of food/water chemicals detected in general). Also, PNS1 score reflects a non-monotone effect for soils (smaller absolute value PNS1 score, higher number of soil chemicals detected).  
PNS2 is largely a contrast between food chemicals (higher PNS2, higher number of food chemicals detected) versus dust/water/soil chemicals (higher PNS2, lower numbers of dust/water/soil chemicals detected in general).
Note that there are strong correlations between groups of chemicals within each type of data (water, dust, food, soil), as the arrows within each chemical group largely follow similar directions (with a few exceptions).
From the PNS scores in the lower plot of Figure \ref{PNS2} we can see that there is a lot of variability, and the higher variance of PNS 1 score in Region A is apparent, as well as the higher PNS score 1 and the higher PNS score 2 on average for Region A.

\subsection{Interpretation}
We now investigate which chemicals have significant differences between regions using Fisher's exact test on detection versus non-detection for Region A versus Region B. 
We use False Discovery Rate FDR=0.1 with \citet{Benjhoch95} correction to give the following chemicals that are significantly less detected in Region A versus Region B:

\begin{small}
\begin{verbatim}                        
Dust: Methyl 1,2,2,6,6-pentamethyl-4-piperidyl sebacate
Dust: Moroctic acid                                    
Dust: p-cresolsulfatepotassium;p-Cresolsulfate         
Dust: alpha-Eleostearic acid                               
Soil: 4-Dodecylbenzenesulfonic acid
\end{verbatim}
\end{small}

In essence, we can see that regional differences are primarily characterized by variation in chemical detection richness, with mainland Miami-Dade environments (Region B) exhibiting more complex chemical mixtures than coastal comparison areas (Region A). 

Possible explanations may be that Region B has more mixed urban land use (traffic, commercial activity, wastewater inputs); greater dust/soil tracking into homes and yard disturbance. These factors increase the number and diversity of detectable organic chemicals in soil and dust.

Region A areas have more coastal sandy soils, which retain fewer organic contaminants; 
more high-rise or hard-floor housing, reducing dust accumulation and soil tracking; and potentially differently sourced water systems. 
Together, these factors could possibly produce lower chemical richness and lower levels of several compounds in Region A, consistent with the patterns observed across dust, and soil. However, there is a lot of household variability in the data and so one must be cautious with interpretation.

In order to investigate the urine chemical associations we carried out 
logistic regression for the binary detection of the chemicals as response, with predictors given by 
the PNS scores 1,2, region, class, sex and age of the children.  There were significant effects for Octanesulfonic acid being detected more frequently in region B, more frequently in Low class, more frequently with higher PNS score 2, and more frequently detected in higher ages (above 30 months). The chemical 4-Dodecylbenzenesulfonic acid was less frequently detected in higher ages (above 30 months), and finally no significant effects were observed for Tripropyl citrate.
Ongoing studies are underway to quantify the concentrations of these chemicals and further evaluate/investigate these observations.

\section{Discussion}
There are many other possible choices of intermediate space for regression with compositional data, including PC scores, Euclidean embeddings, and tangent-space methods. An advantage of the PNS-based intermediate space in our application is that dominant modes of compositional variation are represented by interpretable geometric features, such as circular trajectories corresponding to relative trade-offs among subsets of components, as illustrated in the environmental chemical data. In contrast to log-ratio–based regression approaches, which impose Euclidean structure on the simplex via linearizing transforms, the proposed framework respects the intrinsic curved geometry induced by the Fisher-Rao metric, allowing nonlinear and cyclic modes of variation to be captured more naturally. PNS-based regression is therefore particularly advantageous when compositional variation is concentrated along nonlinear trajectories, for example when relative dominance shifts smoothly between groups of components.

A direct approach to principal subsimplex analysis has been recently developed by \cite{lee2025principalsubsimplexanalysis}, where each subsimplex is formed by combining two components at each stage in a particular ratio. Two approaches are considered: using simplices and using positive orthants. It will be interesting to compare subsimplex analysis using these direct approaches with our PNS method in applications, including the environmental contaminants dataset.

The proposed approach uses a two-stage estimation procedure in which the PNS decomposition is estimated independently of the regression model. While effective in practice, joint estimation of PNS structure and regression parameters may improve efficiency and represents an important direction for future work. Although we focus on the square-root embedding, exploratory analyses with alternative values of $\alpha$ are worth exploring in other applications. It made little practical difference in the 3D application of Section \ref{application} and no difference when compositions are obtained from binarized values. Continued developments in high-dimensional PNS \citep{Monemetal25} further enhance the applicability of the proposed framework to modern compositional datasets.

%for the journal version
%\section*{Data availability}
%The authors plan to make the environmental chemical dataset available online as part of the {\tt shapes} package in R \citep{Rshapes}.

\section*{Acknowledgments}
The authors have no conflicts of interest to declare. Part of the research
was funded by the following award from the Environmental Protection Agency (EPA)
to Natalia Quinete at Florida International University (FIU): EPA-G2020-STAR-D (840199),
Estimating Children’s Soil and Dust Ingestion Rates for Exposure Science. The opinions,
findings, and conclusions presented are of the authors and do not necessarily reflect those
of the EPA. Written consent was obtained from the
families before their participation, following approval by the FIU institutional review board
(IRB–21–0385).

\bibliographystyle{apalike}
\bibliography{fullref-v14}

\begin{thebibliography}{}

\bibitem[Aitchison, 1986]{Aitchison86}
Aitchison, J. (1986).
\newblock {\em The Statistical Analysis of Compositional Data}.
\newblock Chapman and Hall, London.

\bibitem[Aitchison and Bacon-Shone, 1984]{Aitchbaco84}
Aitchison, J. and Bacon-Shone, J. (1984).
\newblock Log contrast models for experiments with mixtures.
\newblock {\em Biometrika}, 71(2):323--330.

\bibitem[Bates, 2005]{Bates05}
Bates, D. (2005).
\newblock Fitting linear mixed models in {R}.
\newblock {\em R News}, 5(1):27--30.

\bibitem[Benjamini and Hochberg, 1995]{Benjhoch95}
Benjamini, Y. and Hochberg, Y. (1995).
\newblock Controlling the false discovery rate: a practical and powerful
  approach to multiple testing.
\newblock {\em J. Roy. Statist. Soc. Ser. B}, 57(1):289--300.

\bibitem[Dryden, 2025]{Rshapes}
Dryden, I.~L. (2025).
\newblock {\em {\tt shapes} package}.
\newblock R Foundation for Statistical Computing, Vienna, Austria.
\newblock Contributed package, Version 1.2.8.

\bibitem[Dryden and Mardia, 2016]{Drydmard16}
Dryden, I.~L. and Mardia, K.~V. (2016).
\newblock {\em Statistical Shape Analysis, with Applications in {R}, 2nd
  edition}.
\newblock Wiley, Chichester.

\bibitem[Fletcher, 2013]{Fletcher13}
Fletcher, P.~T. (2013).
\newblock Geodesic regression and the theory of least squares on {R}iemannian
  manifolds.
\newblock {\em Int. J. Comput. Vis.}, 105(2):171--185.

\bibitem[Gabriel, 1971]{Gabriel1971Biplots}
Gabriel, K.~R. (1971).
\newblock The biplot graphic display of matrices with application to principal
  component analysis.
\newblock {\em Biometrika}, 58:453--467.

\bibitem[Jung et~al., 2012]{Jungetal12}
Jung, S., Dryden, I.~L., and Marron, J.~S. (2012).
\newblock Analysis of principal nested spheres.
\newblock {\em Biometrika}, 99(3):551--568.

\bibitem[Kenward and Roger, 1997]{Kenwroge97}
Kenward, M.~G. and Roger, J.~H. (1997).
\newblock Small sample inference for fixed effects from restricted maximum
  likelihood.
\newblock {\em Biometrics}, 53:983--997.

\bibitem[Lee et~al., 2025]{lee2025principalsubsimplexanalysis}
Lee, H., Hingee, K.~L., Scealy, J.~L., Wood, A. T.~A., Grunsky, E., and Marron,
  J.~S. (2025).
\newblock Principal subsimplex analysis.
\newblock arXiv 2504.09853.

\bibitem[Li et~al., 2023]{Lietal23}
Li, B., Yoon, C., and Ahn, J. (2023).
\newblock Reproducing kernels and new approaches in compositional data
  analysis.
\newblock {\em Journal of Machine Learning Research}, 24(327):1--34.

\bibitem[Mardia and Jupp, 2000]{Mardjupp00}
Mardia, K.~V. and Jupp, P.~E. (2000).
\newblock {\em Directional statistics}.
\newblock Wiley Series in Probability and Statistics. John Wiley \& Sons Ltd.,
  Chichester.

\bibitem[Marron and Dryden, 2021]{Marrdryd22}
Marron, J.~S. and Dryden, I.~L. (2021).
\newblock {\em Object Oriented Data Analysis}.
\newblock CRC Press/Chapman and Hall, Boca Raton.

\bibitem[Monem et~al., 2025]{Monemetal25}
Monem, M., Dryden, I.~L., and George, F. (2025).
\newblock Principal nested spheres for high-dimensional data.
\newblock arXiv 2511.08398.

\bibitem[Ogunbiyi et~al., 2024]{Ogunbiyietal24}
Ogunbiyi, O.~D., Cappelini, L. T.~D., Monem, M., Mejias, E., George, F.,
  Gardinali, P., Bagner, D.~M., and Quinete, N. (2024).
\newblock Innovative non-targeted screening approach using high-resolution mass
  spectrometry for the screening of organic chemicals and identification of
  specific tracers of soil and dust exposure in children.
\newblock {\em Journal of Hazardous Materials}, 469:134025.

\bibitem[Pennec, 2006]{Pennec06}
Pennec, X. (2006).
\newblock Intrinsic statistics on {R}iemannian manifolds: Basic tools for
  geometric measurements.
\newblock {\em Journal of Mathematical Imaging and Vision}, 25(1):127--154.

\bibitem[Scealy and Welsh, 2011]{Scealwels11}
Scealy, J.~L. and Welsh, A.~H. (2011).
\newblock Regression for compositional data by using distributions defined on
  the hypersphere.
\newblock {\em Journal of the Royal Statistical Society: Series B (Statistical
  Methodology)}, 73(3):351--375.

\bibitem[Srivastava et~al., 2007]{Srivastavaetal07}
Srivastava, A., Jermyn, I., and Joshi, S. (2007).
\newblock Riemannian analysis of probability density functions with
  applications in vision.
\newblock In {\em 2007 IEEE Conference on Computer Vision and Pattern
  Recognition}, pages 1--8.

\bibitem[Srivastava and Klassen, 2016]{Srivklas16}
Srivastava, A. and Klassen, E.~P. (2016).
\newblock {\em Functional and Shape Data Analysis}.
\newblock Springer, New York.

\bibitem[van~den Boogaart and Tolosana-Delgado, 2013]{Boogtolo13}
van~den Boogaart, K.~G. and Tolosana-Delgado, R. (2013).
\newblock {\em Analyzing Compositional Data with {R}}.
\newblock Springer, Heidelberg.

\bibitem[{van den Boogaart} et~al., 2024]{Rcomp}
{van den Boogaart}, K.~G., Tolosana-Delgado, R., and Bren, M. (2024).
\newblock {\em compositions: Compositional Data Analysis}.
\newblock R package version 2.0-8.

\end{thebibliography}

\end{document}